\documentclass{article}

\usepackage{PRIMEarxiv}

\usepackage[utf8]{inputenc} % allow utf-8 input
\usepackage[T1]{fontenc}    % use 8-bit T1 fonts
\usepackage{hyperref}       % hyperlinks
\usepackage{url}            % simple URL typesetting
\usepackage{booktabs}       % professional-quality tables
\usepackage{amsfonts}       % blackboard math symbols
\usepackage{nicefrac}       % compact symbols for 1/2, etc.
\usepackage{microtype}      % microtypography
\usepackage{lipsum}
 \usepackage{amsmath}
\usepackage{setspace}
\usepackage{fancyhdr}       % header
\usepackage{graphicx}
\usepackage{indentfirst}
\usepackage{tikz}
\usepackage{listings}
\graphicspath{{media/}}     % organize your images and other figures under media/ folder

\title{A Multi-state Markov Model to Infer the Latent Deterioration Process From the Maintenance Effect on Reliability Engineering of Ships
}

\author{
  Hyunji Moon \\
  IEOR Department \\
  Columbia University, New York  \\
  \texttt{bayes.moon@gmail.com } \\
   \And
  Jungin Choi \\
  Department of Statistics \\
  Seoul National University \\
  \texttt{serimtech07@snu.ac.kr} \\
   \And
  Seoyeon Cha \\
  Department of Plant Science \\
  Seoul National University \\
  \texttt{hh13710@snu.ac.kr} \\
}

\begin{document}
\maketitle

\begin{abstract}

Maintenance optimization of naval ship equipment is crucial in terms of national defense. However, the mixed effect of the maintenance and the pure deterioration processes in the observed data hinders an exact comparison between candidate maintenance policies. That is, the observed data-annual failure counts of naval ships reflect counteracting actions between the maintenance and deterioration. The inference of the latent deteriorating process is needed in advance for choosing an optimal maintenance policy to be carried out. This study proposes a new framework for the separation of the true deterioration effect by predicting it from the current maintenance effect through the multi-state Markov model. Using an annual engine failure count of 99 ships in the Korean navy, we construct the framework consisting of imputation, transition matrix design, optimization, and validation. The hierarchical Gaussian process model is used for the imputation and the three-state Markov model is applied for the estimation of parameters in the deterioration and maintenance effect. To consider the natural (deterioration) and artificial (maintenance) effect respectively, the Bayesian HMM model with a categorical distribution is employed. Computational experiments under multiple settings showed the robustness of the estimated parameters, as well as an accurate recovery of the observed data, thereby confirming the credibility of our model. The framework could further be employed to establish a reliable maintenance system and to reduce an overall maintenance cost. 

\end{abstract}

% keywords can be removed
\keywords{Multi-state Markov model \and equipment reliability and maintenance \and optimization \and imputation}

\section{Introduction}

A maintenance policy is crucial both in terms of the safety and efficiency in managing naval ships’ equipment. Maintenance policy includes several controllable variables to be determined, such as an inspection frequency or an acceptable maintenance standard. Too long inspection interval or overly lenient standard for the repair would result in an unstable system, and the resulting failure costs will be markedly increased. On the other hand, strict maintenance with frequent inspections and excessively conservative standards would yield an excellent budget waste. 

For effective policy development, precise diagnosis of the current system should be first addressed based on the historical failure data. For example, annual engine failure count data is a significant source of information on two main processes: deterioration and maintenance. However, the failure count data ostensibly shows the combined effect; thus, identifying a hidden deteriorating process is required.

Motivated by the fact that engine deterioration is a gradual process and its future depends on the current state, we apply the multi-state Markov model to predict deterioration. The multi-state model is defined as a stochastic process that experiences a few possible states \cite{hougaard99}. This stochastic model could reflect much uncertainty, and the future state could be predicted with randomness. Also, the multi-state approach can provide more interpretability and an effective management policy in the real world since it perceives the whole system as a gradual process between the perfect state (normal) and complete failure (failure) \cite{qiu15, kolobook}. Moreover, computational complexity is significantly reduced due to its discrete state modeling. Markovian multi-state models have been widely used in infrastructure management for the prediction of future deterioration states, such as bridge \cite{kallen07}, healthcare systems \cite{gonzal20}, and pavement \cite{butt87} systems. However, the existing literature mainly focused on risk assessment for a marine domain, and few have been on the failure prediction \cite{roohi14, wang20}. 
Many time-based multi-state Markov approaches assume that the lifetime distribution is known, which generally does not hold \cite{Jonge20}. Since a regular investigation and maintenance have been implemented on each system, most available failure data result from the current management policy. Therefore, deterioration has been ignored in many past studies even though the true deterioration process cannot be seen directly from observed data. Reliable construction of transition probability matrix requires consecutive deterioration data without any maintenance intervention \cite{morcous06}. To deal with this problem, we predict the effect of the latent deterioration process from the current maintenance policy. 

This paper aims to suggest an overall framework for reliability engineering on ships. At the same time, the data is incomplete, and the underlying deterioration process is not directly observed due to the effect of the current maintenance policy. Using the collected annual engine failure count data for the recent ten years from the Korean navy, we estimate the true deterioration process out of maintenance intervention by taking a nonlinear optimization approach. Naval ship equipment failures are assumed as stochastic degradation processes, and they are modeled as three multi-state continuous-time Markov models. The transition probabilities between each deterioration state are represented as a 3×3 transition probability matrix (TPM), which is to be estimated. \cite{li2014deterioration} computed the TPM by constructing the deterioration distribution; however, they lacked the consideration of an age factor. This study estimated the age-dependent transition probabilities with the transition rate matrix obtained by the Kolmogorov equation, an inhomogeneous Markov model. The results have shown different deterioration patterns for each age period.

  \begin{figure}[!htb]
  \centering
  \label{fig:flow}
    \includegraphics[width=1 \textwidth]{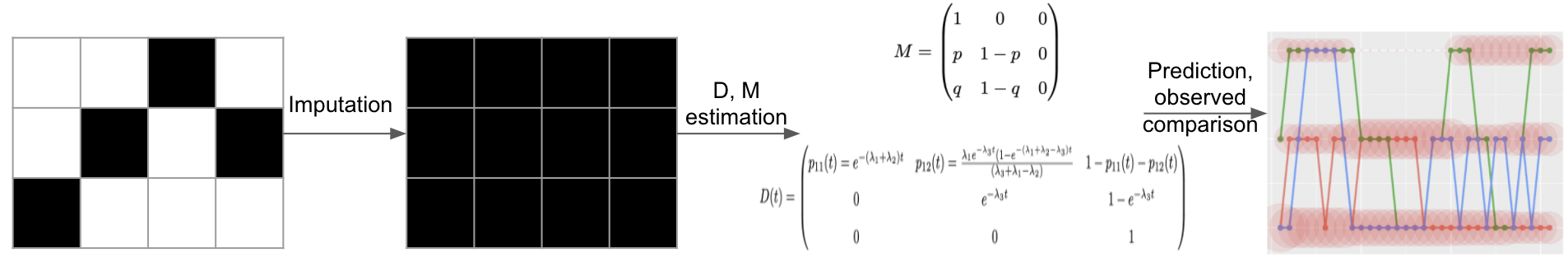}
    \caption{Process of the proposed model.}
  \end{figure}

Figure 1 shows the overall process of our proposed model. First, the missing data is imputed with the Gaussian process model to aid the next step, parameter estimation of transition and maintenance matrix. Without this step, estimation was highly inconsistent. Second, based on the observed data, the best parameter representation of the deterioration and maintenance process was designed. Then these parameters are estimated for given train sets. Lastly, with the resulting parameter values, state series are predicted and compared with real observed states. The last step is necessary for showing the validity of our model. 

The outline of this paper is as follows. In Section 2, we introduce our overall modeling framework and multi-state Markov model. We formally define the deterioration and maintenance matrix in Section 3 and also describe our optimization process to separate the pure effect of the deterioration process and estimate the hidden deterioration rate. Section 4 shows the result of our approach, estimated deterioration parameters, and prediction accuracy. Lastly, limitations and future developments are discussed in Section 5.

\section{Problem setting} 

Annual failure counts of 99 naval ships for the recent ten years (2010-2019) were retrieved from the equipment maintenance information system of the Korean navy. 99 naval ships are classified into five engine types, and Figure 2 presents an overview of our data. The data has 99 vectors with a length of 31, which represents the number of ships and their lifetime. The cells are colored with different colors according to their engine type, and the uncolored cells represent missing data. We arranged the failure data of 99 propulsion ship engines according to their engine types (1 to 5). As we can see, the amount of data for each engine type is highly imbalanced. Considering that the data was recorded in the recent 10 years, the recorded lifetime region differs according to its introduction time. For new models introduced after 2010 (e.g., engine type 5), failure counts were only recorded on younger ages (0-9 yrs), whilst old model’s data introduced in 1985 were constrained on older ages (25-31 yrs). Moreover, a similarity between the data under the same categories could be inferred; for example, same engine types tend to share comparable age ranges and a scale of failure counts. Due to the security problem of military data, only its scaled version is reported throughout the paper.

  \begin{figure}[!htb]
  \centering
  \label{fig:states}
    \includegraphics[width=1 \textwidth]{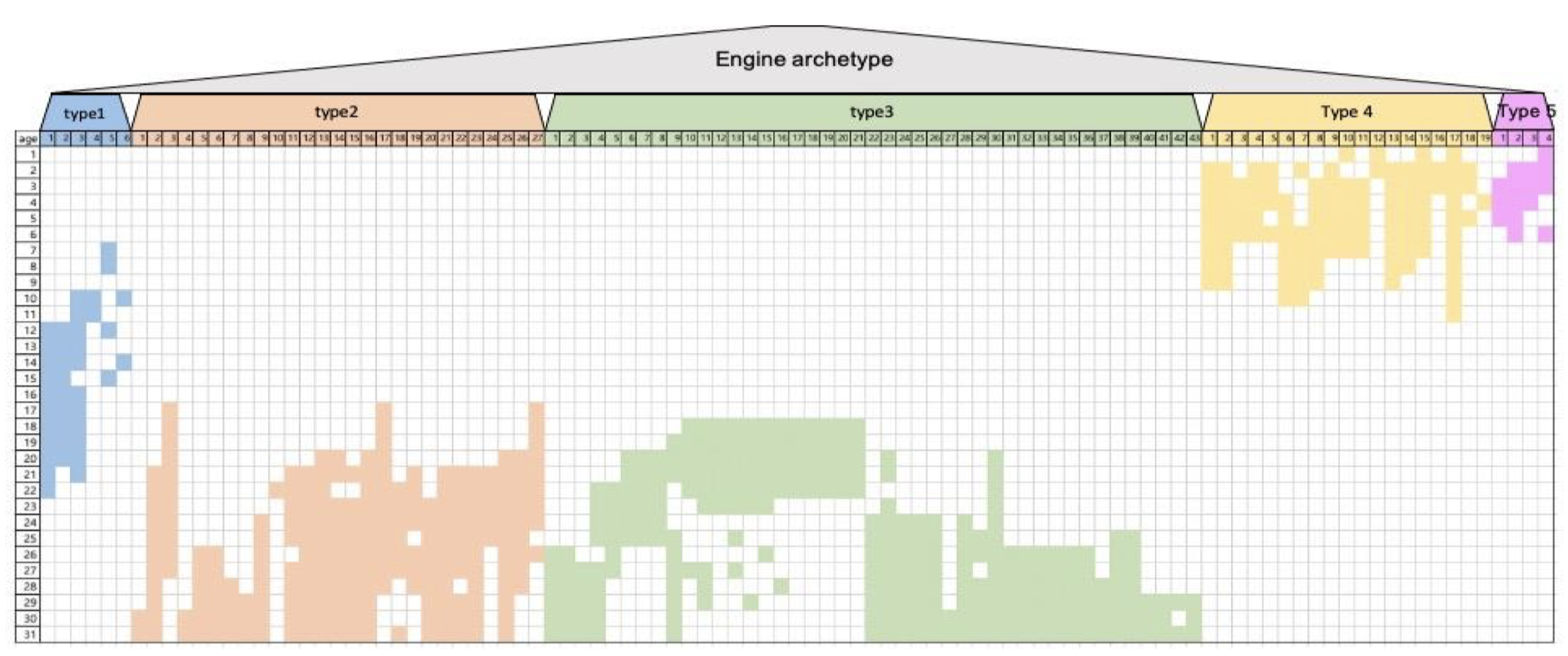}
    \caption{Overview of the failure counts from the 99 Korean naval ships.}
    \label{fig:GSD}
  \end{figure}

\section{Imputation using a Gaussian Process}

The original failure counts of 99 naval ships are largely missing and unevenly spaced, imputation is required to achieve a robust multi-state Markov model. After comparing several models including a hierarchical spline model and a Poisson process, a hierarchical Gaussian process model was chosen for the imputation. The hierarchical structure of the engine types in the system validates the effectiveness of the applied hierarchical model \cite{gelman2013bayesian}.

% It is notable that due to the hierarchical structure of a system, such as shared engine types, using a hierarchical model is effective 

We assume that failure counts follow a normal distribution with mean $\mu_{t,j}$ and variance $\sigma^2_{k,j}$ for $j$th ship and time $t$, where $k[j]$ is engine index for $j$th ship.
This is engine index for jth ship. This is based on the observation that the variances of failure counts between each engine type look quite distinct. For example, while engine types 4 and 5 show a large variance, type 3 shows a small variance of the given data.

\begin{equation} \label{eq1}
y_{t,j} \sim \mathcal{N}(\mu_{t,j},\sigma^2_{k[j]})
\end{equation}

Mean $\mu_{t,j}$ is defined as an additive Gaussian process which is the sum of variables with normal distribution and Gaussian processes.
$\mu$ is an overall mean of $\mu_{t,j}$, and $\theta_t^{age}, \theta_j^{ship}, and \theta_{k[j]}^{engine}$ follow a normal distribution with zero mean and variance $\sigma^2_{age}, \sigma^2_{ship}, and \sigma^2_{engine}$, respectively.

\begin{equation} \label{eq2}
\begin{split}
\mu_{t,j} &= \mu + \theta_t^{\ age}+\theta_j^{\ ship}+\theta_{k[j]}^{\ engine}+\gamma_{t,j}+\delta_{t,\ k[j]} \\
t &=1,2 \dots T, \ \  j =1,2, \dots N
\\
\end{split}
\end{equation}

\begin{equation} \label{eq3}
\begin{split}
\theta_t^{\ age} &\sim \mathcal{N}(0, \sigma^2_{age}\ )
\\
\theta_j^{\ ship} &\sim \mathcal{N}(0,\sigma^2_{ship}\ )
\\
\theta_k^{\ engine} &\sim \mathcal{N}(0,\sigma^2_{engine}\ )
\end{split}
\end{equation}

\begin{equation}
\label{eqn:eq4}
\begin{split}
\gamma_{j} &\sim \mathcal{N}(\textbf{0},\pmb{K}_{l^{\gamma},{\alpha}^{\gamma}})
\\
\delta_{k} &\sim \mathcal{N}(\textbf{0},\pmb{K}_{l^{\delta},{\alpha}^{\delta}})
\end{split}
\end{equation}

$\gamma_j$ and $\delta_k$ are Gaussian processes whose $t$th elements are $\gamma_{t,j}$ and $\delta_{t,k[j]}$, respectively.
The covariance kernel for the Gaussian processes $\gamma_{j}$ and $\delta_{k}$ are defined as an exponentiated quadratic kernel. The exponentiated quadratic kernel defines the covariance of each Gaussian process between $f(x_i)$ and $f(x_j)$ where  $f:\mathbb{R}^D \rightarrow \mathbb{R}$ as a function of the squared Euclidian distance between  $x_i \in \mathbb{R}^D$ and $x_j \in \mathbb{R}^D$ : 

\begin{equation}
\label{eqn:eq5}
\begin{split}
\text{Cov}(f(x_i),f(x_j))&=k(x_i,x_j) \\
&= \alpha^2 \exp \left( -\frac{1}{2l^2} \sum_{d=1}^{D}(x_{i,d}-x_{j,d})^2 \right)
\end{split}
\end{equation}

with $\alpha$ and $l$ constrained to be positive.

Since the covariance function $\pmb{K}_{l^{\gamma},{\alpha}^{\gamma}}$ and $\pmb{K}_{l^{\delta},{\alpha}^{\delta}}$ follow an exponentiated quadratic kernel, $ij$th element of $\pmb{K}_{l^{\gamma},{\alpha}^{\gamma}}$ and $\pmb{K}_{l^{\delta},{\alpha}^{\delta}}$ are defined as \ref{eqn:eq6}

\begin{equation}
\label{eqn:eq6}
\begin{split}
(\pmb{K}_{l^{\gamma},{\alpha}^{\gamma}})_{ij} &=  {{\alpha}^{\gamma}}^2 \exp \left(-\frac{1}{2{l^{\gamma}}^2} \sum_{d=1}^{D}(x_i,d-x_j,d)^2 \right) \\
(\pmb{K}_{l^{\delta},{\alpha}^{\delta}})_{ij} &=  {{\alpha}^{\delta}}^2 \exp\left(-\frac{1}{2{l^{\delta}}^2} \sum_{d=1}^{D}(x_i,d-x_j,d)^2\right)
\end{split}
\end{equation}

The prior distribution for $l^{\gamma}$ and $l^{\delta}$ is the Weibull distribution with shape parameter $k$ and scale parameter $\lambda$. Since their mean largely depends on the scale parameter, we reparametrized $l^{\gamma}$ and $l^{\delta}$ so that the scale parameter $\lambda$ is fixed to 1.

\begin{equation} \label{eq7}
\begin{split}
l_{\gamma} &\sim \text{Weibull}(k_{\gamma},\lambda_{\gamma}) 
\\
l_{\delta} &\sim \text{Weibull}(k_{\delta},\lambda_{\delta}) 
\end{split}
\end{equation}

are reparametrized into 

\begin{equation} \label{eq8}
\begin{split}
l_{\gamma}^{\ s} &\sim \text{Weibull}(k_{\gamma},1) ,\ \ \ \ l_{\gamma} = l_{\gamma}^s * \lambda_{\gamma}
\\
l_{\delta}^{\ s} &\sim \text{Weibull}(k_{\delta},1) ,\ \ \ \ l_{\delta} = l_{\delta}^s * \lambda_{\delta}
\end{split}
\end{equation}

Using this model, we gained an appropriate posterior distribution for each parameter of the Gaussian model, thus imputing the missing data using the best results in terms of prediction accuracy.

\section{Separating the Deterioration and Maintenencnce effect using CTMC}

\subsection{Continuous time multi-state Markov model}

In this study, we propose a three-state continuous-time Markov model to describe the deterioration states over the naval ship’s life cycle as Figure \ref{fig:states}. The ship’s deterioration state could be classified into three states according to their annual failure counts by using a quantile classification method as shown in Table \ref{tbl:my-table}. The corresponding failure counts are standardized by a usual procedure, subtracting the mean and dividing by its standard error. Each deterioration state of engines has a range from 1 to 3, where 1 corresponds to a "normal" state, 2 corresponds to a "near failure" state, and 3 to a complete "failure" state of engines. The state of each ship at time t is assigned by the standardized annual failure counts as shown in Table 1.

\begin{figure}[!htb]
    \centering
        \includegraphics[width=0.4 \textwidth]{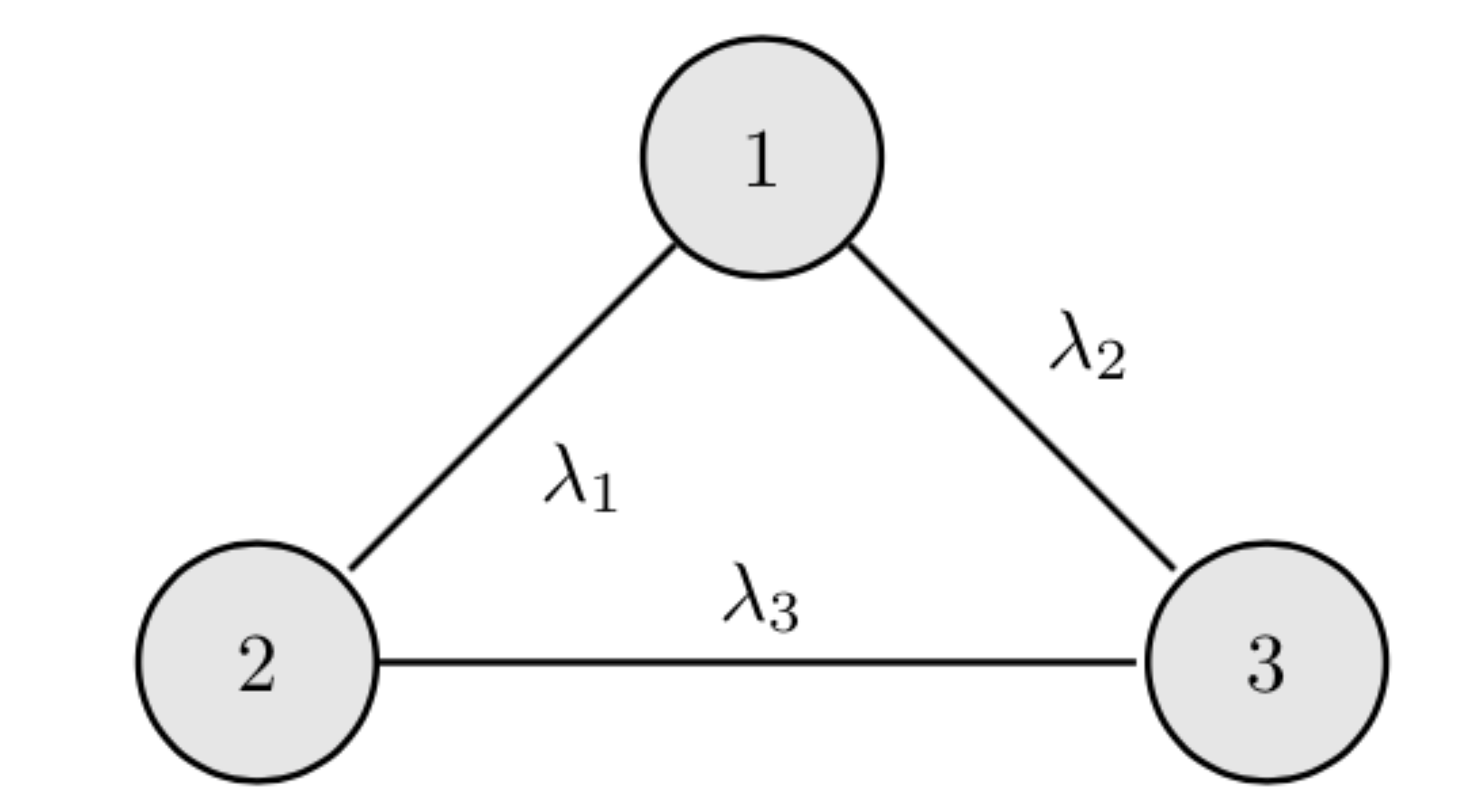}
\caption{Deterioration rate between three states.}
\label{fig:states}
\end{figure}

\begin{table}[h]
\caption{Deterioration state of ship equipment according to the standardized annual failure counts.}
\centering
\begin{tabular}{ccc}
\hline 
\text{State} & \text{Status Description} & \text{Annual failure counts} \\ \hline
\text{1 } & \text{Normal}           & [ -1.8304, -0.3340 )           \\ 
\text{2 } & \text{Near Failure}         & [ -0.3340, -0.0703 )      \\ 
\text{3 } & \text{Failure}    &  [ -0.0703,  2.1273 ] \\ \hline
\end{tabular}

\label{tbl:my-table}
\end{table}

A continuous-time multi-state model is for a continuous-time stochastic process where individuals can occupy a finite number of states. 
Let $\left\{Y(t) | t \in T \right\}$ represent the deterioration state of the process at time $t$. The transition between states follows a continuous-time Markov process, and transition probabilities only depend on a present state. 
For a time-homogeneous Markov chain, we can write the transition probability function from state $i$ at time $s$ to state $j$ at time $t$ as $p_{ij}(s, t) = Pr\left\{Y(t) = j | Y(s) = i\right\}$. 
The transition probabilities between all possible pairs $(i, j)$ are represented by a $n \times n$ matrix called the transition probability matrix $P(s, t)$, where $n$ is the number of possible condition states as shown in \ref{eq9}.

\begin{equation}
\label{eq9}
P(s, t) = \begin{pmatrix}
p_{11}(s, t) & p_{12}(s, t) & p_{13}(s, t)\\
p_{21}(s, t) & p_{22}(s, t) & p_{23}(s, t) \\
p_{31}(s, t)& p_{32}(s, t)& p_{33}(s, t)
\end{pmatrix}
\end{equation}

In a continuous-time Markov process, transition rate matrix $Q$ is also introduced to explain instantaneous transition rates between states. 
The transition rate matrix $Q(t)$, also known as an intensity matrix, has an element $q_{ij}$ (for $i \neq j$) denoting the rate departing from $i$ and arriving in state $j$. Diagonal elements $q_{ii}$ are defined such that ${\displaystyle q_{ii}=-\sum _{j\neq i}q_{ij}.}$, satisfying the row-wise sums of the matrix to zero. 
In other words, the off-diagonal elements of $Q$ represent the rates governing the exponentially distributed variables that are used to describe the amount of time that elapses before a particular type of base substitution occurs.

Transition matrix $P(t)$ can be calculated using the Kolmogorov's forward and backward equations, $P'(t) = P(t)Q$ and $P'(t) = QP(t)$, respectively.
The solution is uniquely derived in terms of the transition rate matrix $Q$ as \ref{eqn:eq10} by the matrix exponential of $Qt$.

\begin{equation}
\label{eqn:eq10}
\begin{split}
P(t) &= \exp(Qt)\\
&= \sum_{r = 0} ^{\infty} Q^{r} \frac{t^r}{r!}
\end{split}
\end{equation}

\subsection{Deterioration and maintenance matrix}

To separate the effect of the deterioration and the maintenance, we first define a deterioration matrix $D(t)$, which is a function at the age $t$. Deterioration matrix $D(t)$ should be an upper-diagonal matrix reflecting the reality that engines continuously undergo degradation through the life cycle. 
Also, since the extent of the deterioration varies between different age ranges, it is constructed in a time-inhomogeneous manner. 
Older engines are prone to be readily weakened, and it's reasonable to design a more severe deterioration matrix $D(t)$ for them. 
Equation \ref{eq11} shows the deterioration rate matrix $Q(t)$ for the three-state Markov model. 
This transition rate matrix requires three distinct parameters for each time $t$, and is also upper-diagonal, considering that the deterioration process can only occur in a forward direction. 

\begin{equation}
\label{eq11}
Q(t) = \begin{pmatrix}
-(\lambda_{t,1} + \lambda_{t,2}) & \lambda_{t,1} & \lambda_{t,2}\\
0 & -\lambda_{t,3} & \lambda_{t,3} \\
0 & 0 & 0 \\
\end{pmatrix}
\end{equation}

As in equation 12, transition probabilities of the \textit{deterioration matrix} $D(t)$ can be calculated solely with the given rate matrix $Q(t)$, which comes from the solution of Kolmogorov’s forward and backward equation. Each element in 3 × 3 deterioration matrix $D(t)$ is the deterioration probability, and $D_{ij}(t)$ is the probability of transitioning from state $i$ to $j$.

\begin{equation}\label{eq12}
\begin{split}
D(t) &= \exp(Qt)\\
&= \sum_{r = 0} ^{\infty} Q^{r} \frac{t^r}{r!}\\
&= I + \sum_{r = 1} ^{\infty} Q^{r} \frac{t^r}{r!}
\end{split}
\end{equation}

The resulting exponential matrix can be expressed in a closed form, because the deterioration matrix $D(t)$ is positive-definite, which makes possible for $Q$ to be always factorized as $Q = UDU^{-1}$ using eigen-decomposition, where $D$ is the diagonal matrix of eigenvalues and $U$ is the matrix whose columns are the corresponding eigenvectors. 
Using the diagonal matrix $D$, the deterioration matrix $D(t)$ for three-state Markov model is computed as \cite{jones17}. 
$p_{ij}(t)$ is equal to the occurence probability from state $i$ to $j$.

\begin{equation}
\label{eq13}
D(t) = \begin{pmatrix}
p_{11}(t) = e^{-(\lambda_{1} + \lambda_{3})t} & 
p_{12}(t) = \frac{\lambda_{1} e^{-\lambda_{3}t} (1- e^{-(\lambda_{1} + \lambda_{2} -\lambda_{3})t})}{(\lambda_{1} + \lambda_{2} - \lambda_{3})}& 1 - p_{11}(t) - p_{12}(t) \\
0 & e^{-\lambda_{3}t} & 1 - e^{-\lambda_{3}t} \\
0 & 0 & 1
\end{pmatrix}
\end{equation}

In the same way, we define a \textit{maintenance matrix} $M$ as in \ref{eq14}, which is a transition probability matrix with two parameters. The maintenance matrix $M$ is multiplied to the state probability vector whenever the maintenance is performed. The construction of this maintenance matrix is based on several circumstantial assumptions as follows. First, maintenance interval is once a year. Second, an imperfect maintenance makes the transition probability from state 2 to be divided into state 1 and 2, which is $p_{21}$ and $1-p_{21}$ for certain probability. Likewise, the transition from state 3 could be parameterized with $p_{31}$, $p_{32}$, and $1-p_{31}-p_{32}$. However, the preliminary experiment results showed that $1-p_{31}-p_{32}$ term mostly converged to 0, we assumed it as 0 and replaced the probability $p_{31}$, $p_{32}$ with $p_{31}$ and $1-p_{31}$, discarding the redundant parameter $p_{32}$.

\begin{equation}
\label{eq14}
M=
\begin{pmatrix}
1 & 0 & 0\\
p_{21} & 1-p_{21} & 0\\
p_{31} & 1-p_{31} & 0
\end{pmatrix}
\end{equation}

\subsection{HMM for the estimation of deterioration matrix}

The main objective of our paper is to infer the latent deterioration effect from the observed data intervened by the maintenance effect. This can be achieved by estimating the underlying deterioration probabilities in the deterioration matrix $D(t)$, which could be parametrized by the deterioration rate parameters. The estimation of deterioration rate parameters $\lambda$ remains to be the main challenge.

For the parameter estimation, we consider a hidden Markov model (HMM) with the assumption that observed failure count $y_t$ at time $t$ is generated based on the hidden deterioration state vector \\ $d(t) = (d_{t,1}, d_{t,2}, d_{t,3})^\top$, where $d_{t,j}$ is the probability of state $j$ at time $t$, hence a 3-simplex vector. According to HMM, the output of failure count $y_t$ follows a categorical distribution, $y_t \sim \textit{categorical} (d(t))$, which reflects stochastic randomness of nature. 

With respect to the Markov property, the present deterioration state vector $d(t)$ only depends on the past vector $d(t-1)$. Starting from the initial deterioration state vector, $d(0) = (1,0,0)^\top$, the ships continuously undergo the deterioration process and the maintenance simultaneously at each time, which is modeled with the matrix $D(t)$ and $M$ as mentioned. Taken together, the deterioration state vector at time $t$ is formulated as below:

\begin{equation*}
\label{eq15}
d(t) = D(t) \{ \prod_{i=1} ^{t-1}  M D(i) \} d(0), \ \
y_t \sim  \textit{categorical} (d(t))
\end{equation*}

We took a Bayesian paradigm for the estimation of established parameters with the probabilistic statistical programming language, Stan, for a flexible Bayesian statistical analysis \cite{carpenter2017stan}. Stan takes advantage of the Hamiltonian Monte Carlo (HMC) algorithm to implement the Markov chain Monte Carlo (MCMC) sampling, which is very efficient and powerful to estimate the posterior distribution and where we are struggling to attain. 

In our problem, the target parameters are $\lambda_1$, $\lambda_2$, $\lambda_3$, $p_{21}$, and $p_{31}$, hereafter denoted $\theta$ all at once, and the likelihood function $p(y|\theta)$ is a categorical distribution. As no specific priors are assigned in our code, Stan automatically used a uniform prior to yielding a final posterior distribution $p(\theta|y)$. Stan code to fit the aforementioned model is illustrated in detail below. 

\begin{verbatim}
data {
  int<lower=0> N; // numbmer of obs
  int<lower=0> T; // number of time bins = max age
  int<lower=1>S; // number of states
  int<lower=1> P; // number of periods for inhomogenous dtr.
  int<lower=0, upper =S> states[N]; //observed state
  int obs2time[N]; // map obs to time
  int<lower=0, upper=S> initial_state;
}

transformed data {
  vector[S] initial;
  vector<lower=0>[S] alpha;
  vector<lower=0>[S] beta;
  initial = rep_vector(0, S);
  initial[initial_state] = 1;
  alpha = rep_vector(3,3);
  beta = rep_vector(3,3);
}

parameters {
  real<lower=0> rate[S];
  real<lower=0, upper=1> p21;
  real<lower=0, upper=1> p31;
}

transformed parameters {
  matrix[S, S] Dtr; // Deterioration
  matrix[S, S] Mnt; // Maintenance
  simplex[S] latent_states[T];
  matrix[S, S] tmp_p;
  // Maintenance
  // is left stoch. matrix, transpose of eq.14 from the paper
  Mnt = [[1, p21, p31],
         [0, 1-p21, 1-p31],
         [0, 0, 0]];
  // Deterioration by period
  // is left stoch. matrix, transpose of eq.13 from the paper
  tmp_p[1,1] = exp(-rate[1]- rate[2]);
  tmp_p[2,1] = rate[1] * exp(-rate[3]) * (1-exp(-(rate[1]+ rate[2] - rate[3]))) / 
              (rate[1]+ rate[2] - rate[3]);
  tmp_p[3,1] = exp(-rate[3]);
  Dtr = [[tmp_p[1,1], 0, 0],
            [tmp_p[2,1], tmp_p[3,1], 0],
            [1 - tmp_p[1,1] - tmp_p[2,1], 1 - tmp_p[3,1], 1]];
  // Inhomogenous Dtr
  latent_states[1] = Dtr * initial;
  for (t in 2:T){
    latent_states[t] =  (Dtr * Mnt) *latent_states[t-1]; 
    //matrix_power((Dtr * Mnt), (t-1)) * initial;
  }
}

model {
  for (n in 1:N){
    states[n] ~ categorical(latent_states[obs2time[n]]);
  }
}
\end{verbatim}

\section{Result}

\subsection{Estimated parameters}
As we parameterized deterioration probabilities and the maintenance probabilities, the deterioration matrix $D(t)$ is parameterized by the rate matrix $Q$ including 3 rate parameters for each 4 period, totally 12 deterioration rate parameters. The maintenance matrix $M$ is parameterized by 2 parameters, $p$ and $q$. With the observed deterioration data, we split the total data into training and test sets by 94:5 ratio for 1000 different combinations. 

\begin{figure}[!htb]
    \centering
        \includegraphics[width=0.95 \textwidth]{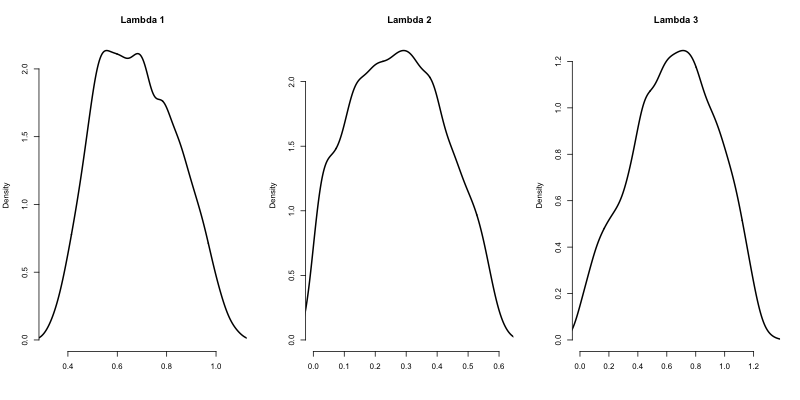}
        \caption{Resulting probability density graph of rate parameters: $\lambda_1, \lambda_2$, and $\lambda_3$.}
        \label{fig:lambda}
\end{figure}

Table \ref{tbl:lambda} shows the estimated average values and standard deviation of rate parameters $\lambda_1$, $\lambda_2$, and $\lambda_3$ . On the whole, $\lambda_1$ and $\lambda_3$ are quite larger than $\lambda_2$, implying that transition from state 1 to 2, and 2 to 3 is more likely than from state 1 to 3. Figure \ref{fig:lambda} shows the probability density graph of the three rate parameters: $\lambda_1, \lambda_2$, and $\lambda_3$, displaying similar shapes between each parameter with certain peak values.

\begin{table}
\caption{Estimated mean value and their standard deviation of rate parameters: $\lambda_1$, $\lambda_2$, and $\lambda_3$} % title name of the table  
  \centering
  \begin{tabular}{lll}
    \toprule
    $\lambda_i$ & Mean & SD  \\ 
    \midrule
    $\lambda_1$& 0.679 &  0.161 \\
    $\lambda_2$& 0.274 &  0.149 \\
    $\lambda_3$& 0.649 &  0.287 \\
    \bottomrule
  \end{tabular}
\label{tbl:lambda}
\end{table}

\begin{figure}[!htb]
    \centering
        \includegraphics[width=0.8 \textwidth]{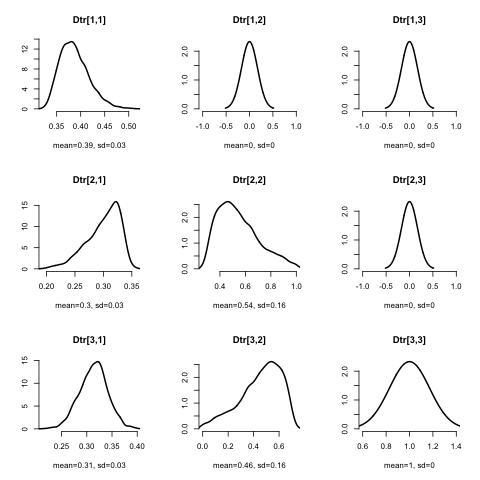}
        \caption{Estimated probability density of the deterioration matrix $D(t)$.}
        \label{fig:deter}
\end{figure}

Figure \ref{fig:deter} shows the estimated distribution of every component of the deterioration matrix $D(t)$. Since $D_{(t)}$ expresses the transition probabilities between 3 deterioration states, $D_{12}(t)$, $D_{13}(t)$ and $D_{23}(t)$ are set to zero, and $D_{33}(t)$ are set to 1 for every period. Each element of $D(t)$ has distinctive peaks on its probability density. Table \ref{tbl:pq} shows the estimated mean value and their standard deviation of components of the deterioration matrix $D_{(t)}$. The deterioration probability from state 1 to 1,2, and 3 shows almost equal probability around 0.3. Conversely, the transition probability from state 2 to 3 is almost 0.6, while the staying probability on state 2 is almost half of it. This implies that the deteriorating process is highly likely from state 2.

\begin{table}
    \caption{Estimated average value and standard deviation of deterioration matrix: $D(t)$.} % title name of the table  
  \centering
  \begin{tabular}{lll}
    \toprule
    $D_{ij}$ & Mean & SD \\ 
    \midrule
    $D_{11}$& 0.387 & 0.0302 \\
    $D_{12}$& 0 & 0 \\
    $D_{13}$& 0 & 0 \\
    $D_{21}$& 0.300 & 0.0294 \\
    $D_{22}$& 0.545 & 0.1603 \\
    $D_{23}$& 0 & 0 \\
    $D_{31}$& 0.313 & 0.0280 \\
    $D_{32}$& 0.455 & 0.1603 \\
    $D_{33}$& 1 & 0 \\
    \bottomrule
  \end{tabular}
  \label{tbl:deter}
\end{table}

\begin{figure}[!htb]
    \centering
        \includegraphics[width=0.8 \textwidth]{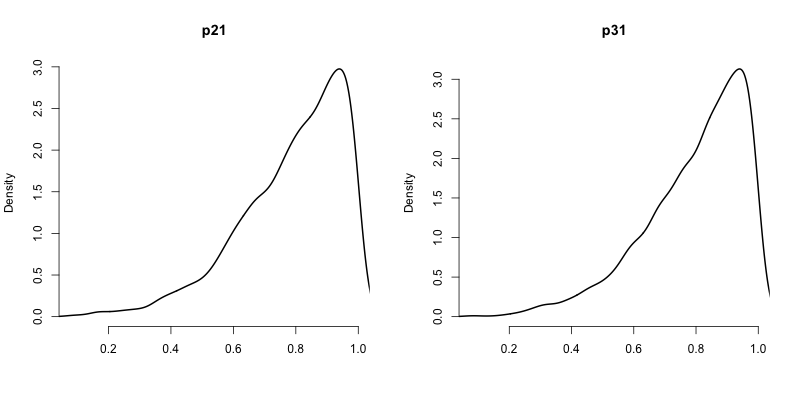}
        \caption{Resulting sampling results of the maintenance parameters: $p_{21}$ and $p_{31}$.}
    \label{fig:pq}
\end{figure}

On the other hand, for the maintenance parameters, $p21$ and $p31$ showed an average value of 0.787 and 0.794, respectively. Table \ref{tbl:pq} shows the estimated average values and standard deviation of rate parameters $p_{21}$ and $p_{31}$. This indicates the transition probability from state 2 to 1 and from state 3 to 2 after the yearly maintenance, implying that engine's states tend to go to state 1 with high probability for both state 2 and 3 engines after the yearly maintenance. Thus, we observed the fine quality of the current maintenance and repair of the Korean navy. Similarly, figure \ref{fig:pq} shows the resulting sampling results of two maintenance parameters, $p_{21}$ and $p_{31}$.

\begin{table}
 \caption{Estimated mean value and their standard deviation of maintenance parameters: $p_{21}$ and $p_{31}$}
  \centering
  \begin{tabular}{lll}
    \toprule
    $p_{ij}$ & Mean & SD  \\
    \midrule
    $p_{21}$& 0.787 & 0.167 \\
    $p_{31}$& 0.794 & 0.167 \\
    \bottomrule
  \end{tabular}
  \label{tbl:pq}
\end{table}

\begin{figure}[!htb]
    \centering
        \includegraphics[width=0.8 \textwidth]{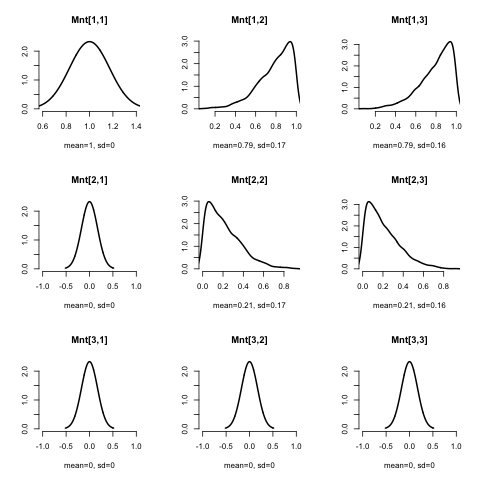}
        \caption{Estimated probability density of the maintenance matrix $M$.}
    \label{fig:mainten}
\end{figure}

According to the estimated values of $p_{21}$ and $p_{31}$, Figure \ref{fig:mainten} shows the resulting distribution of every component of the maintenance matrix $M$. Since $M$ is composed of maintenance probabilities, $M_{12}$,$M_{13}$,$M_{23}$ and $M_{33}$ are fixed to 0 and $M_{11}$ is fixed to 1.

\clearpage

\subsection{Predicted states} 

\begin{figure}[!htb]
  \centering
    \includegraphics[width=1 \textwidth]{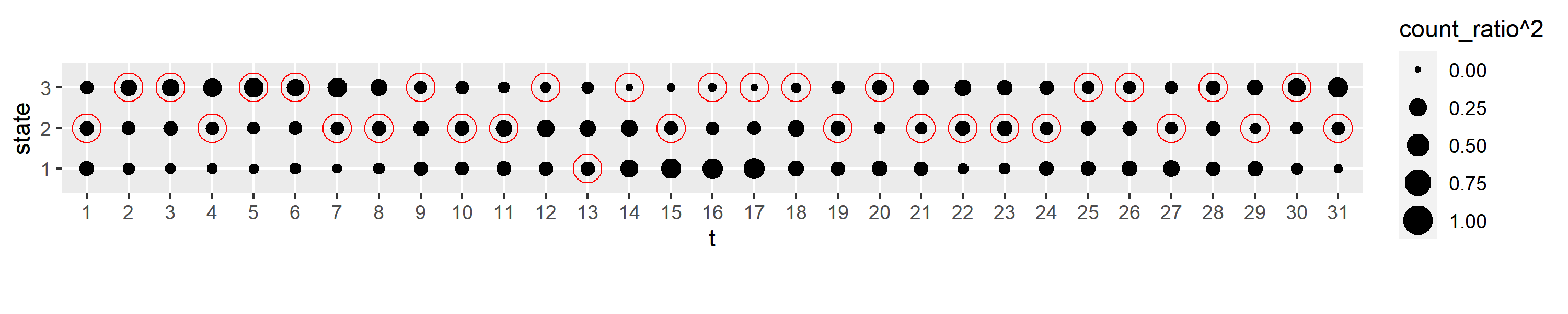}
    \caption{Predicted states and observed states over time. Size of point reflects ratio of states over 99 ships. }
  \label{fig:pred}
\end{figure}

The main purpose of the model is to predict the deteriorating state of each engine so that management could be calculated based on their predictions. Figure \ref{fig:pred} shows the observed ratio of the 99 naval ship's deterioration states, representing their observed probabilities with the size of the black circles.
%i.e. count ratio of each state over 99 ships is reflected in the size of the black circles. 
Red circles indicate the predicted states for each time. Predictions fit well with the observed state from three out of the sample ship's engines in most cases. For example, between years 10 and 20, the flat area which corresponds to the flat area of a bathtub in previous studies, the observed states are concentrated in the first two states and so are the predictions. 

One of the best ways to summarize the accuracy of the prediction is by measuring the distance between the observed and predicted states. Equation \ref{eqn:MSE} shows the error measure between two state series, where $N$ denotes the number of predicted ships, $d_{i}(t)$ and $\hat{d}_{i}(t)$ is the observed and predicted deterioration state of $i$th ship at time $t$, respectively. Note that the value of N is 5 for the test set and 94 for the training set. 

Figure \ref{eqn:MSE} shows the histogram of MSE values between the observed and predicted states for each training and test set. A total of 1000 tests have been repeated and test sets are constructed by randomly selected five engines. Though the training error (average 20.7) is smaller than the test error (average 20.9), MSE values of both training and test sets were barely different, thus confirming the validity of our multi-state Markov model with HMM.

\begin{equation}
 MSE =  \frac{1}{N} \frac{1}{31}\sum_{i=1}^{N}\sum_{t=1}^{31} (d_{i}(t) - \hat{d}_{i}(t))^2 
 \label{eqn:MSE}
\end{equation}

\begin{figure}[!htb]
  \centering
    \includegraphics[width=0.8 \textwidth]{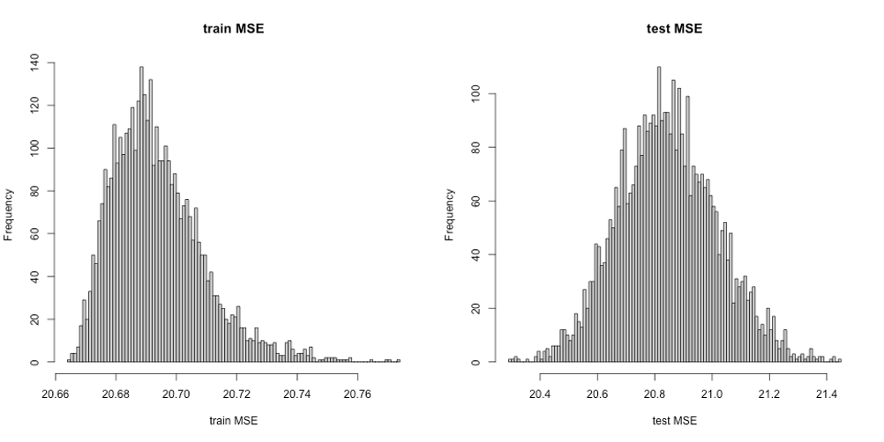}
    \caption{Histogram of the training and test sets MSE values.}
    \label{fig:MSE}
\end{figure}

\subsection{Simulation Based Calibration: SBC} 

Since we estimated the rate parameters in the deterioration matrix $D(t)$, and maintenance parameters in the maintenance matrix $M$, we should validate whether the MCMC computation is correct or not. Since our model is complicated in implementations and algorithms, we validated inferences from Bayesian algorithms capable of generating posterior samples through Simulation-Based Calibration: SBC\cite{Stalts18}. 

\begin{figure}[!htb]
\label{fig:SBC}
  \centering
    \includegraphics[width=0.8 \textwidth]{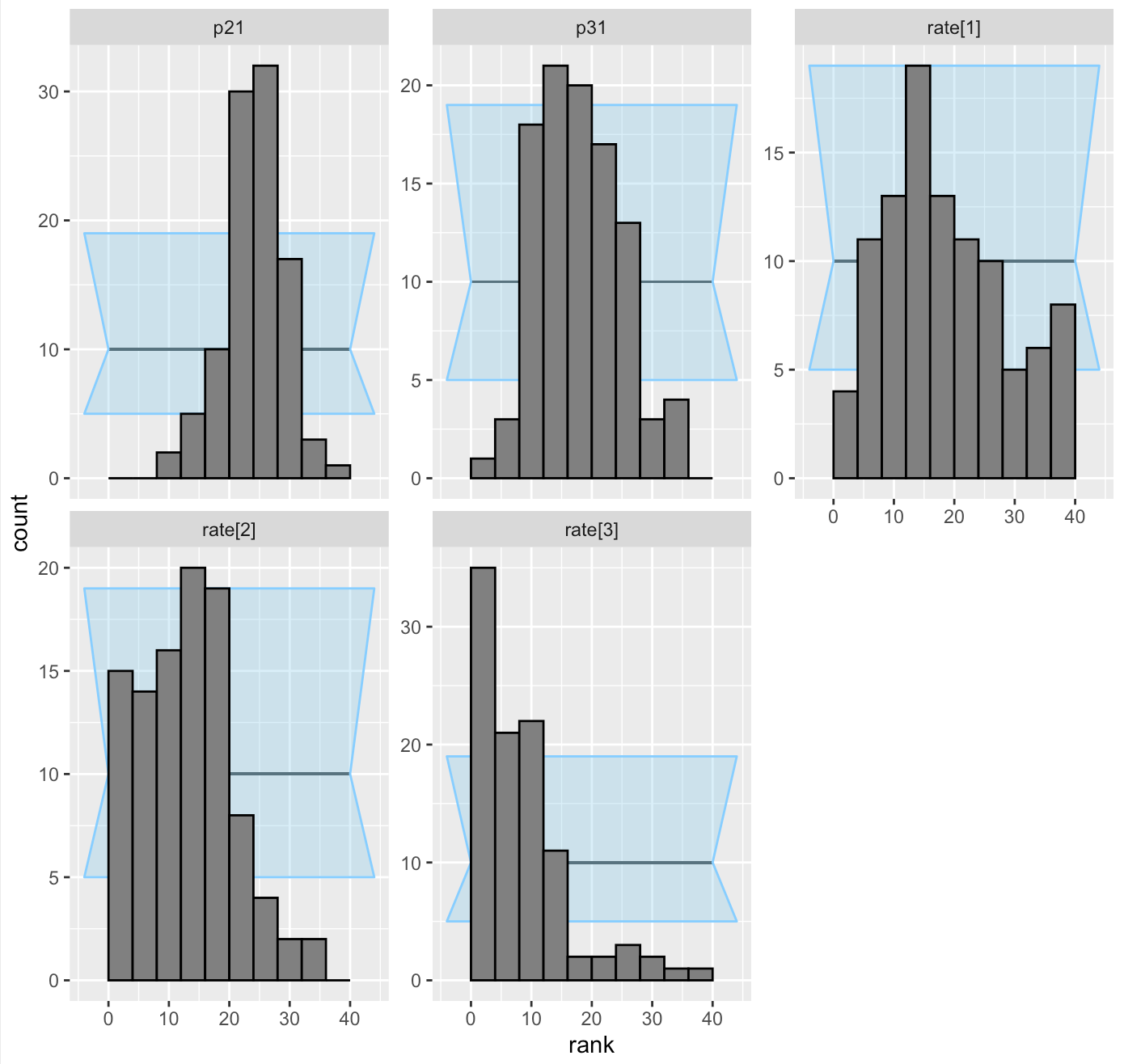}
    \caption{The SBC histogram for each model parameter: $p_{21}$, $p_{31}$, $\lambda_1$, $\lambda_2$, and $\lambda_3$.}
\end{figure}

Figure \ref{fig:SBC} shows the SBC histograms for each model parameter which are the maintenance parameters $p_{21}$, $p_{31}$, and rate parameters $\lambda_1$, $\lambda_2$, and $\lambda_3$. Since SBC histograms for $p_{31}$, $\lambda_1$, and $\lambda_2$ are fairly uniformly distributed and their distributions are mostly included in the blue zone, which identifies accurate computation and consistencies in model implementations. However, SBC histrograms for $p_{21}$ and $\lambda_3$ are not uniformly distributed that infers inaccurate computation. 

Graphical summaries in Figure \ref{fig:SBC} also the nature of the problems of computations in $p_{21}$ and $\lambda_3$\cite{Stalts18}.

Since the SBC histogram for $p_{21}$ is $\cap$-shaped, it indicates that the computed posterior distribution is over-dispersed relative to the prior distribution. This implies that on average, the computed $p_{21}$ posterior will be wider than the true $p_{21}$ posterior. Since $p_{21}$ reflects maintenance from State 2 to 1, the true maintenance effect will be narrower than the computed posterior distribution. 

Since the SBC histogram for $\lambda_3$ is right-skewed, it indicates that the algorithm is strongly biased towards larger values of $\lambda_3$ in the true posterior. This might be caused by the identifiability of $\lambda_3$ and the heterogeneous Markov chain. Since $\lambda_3$ reflects deterioration from State 2 to 3, the algorithm is biased on stronger deterioration from 'Near Failure' to 'Failure' status in the true posterior.

\section{Conclusion}

The multi-state model has a strong potential in reliability analysis. We have applied this framework to the Korean Naval ships’ engine failure data and established a three-state Markov model on the engine deterioration process. The difficulty in maintenance policy suggestion is the inference of the pure deterioration matrix since the current implemented policy is reflected in the given data. We separately estimated the deterioration and maintenance effect using nonlinear optimization based on the imputed data using the hierarchical Gaussian process. Our approach provides a basis for a reasonable policy comparison. We expect maintenance policy suggestions based on our methodology to be the main direction for further research.

The limitation or further development of this study is as follows. First, the deterioration matrix D(t) could be learned hierarchically. Currently, D(t) of each engine type (1 to 5) is learned in a no-pooling way but it could be updated to partial pooling, in other words, in a hierarchical way. Moreover, the rate between four separated time regions in each ship could be partially pooled as well. Though parameters would be added from this additional structure, as has been introduced in \cite{mcelreath2020statistical} and \cite{gelman2013bayesian}, a hierarchical model with its advanced flexibility could increase the model’s accuracy. Second, prior knowledge of the engines (deterioration and operation patterns) could be reflected. Priors could greatly affect the performance of the model if properly understood in the context of the entire Bayesian analysis, from inference to prediction to model evaluation \cite{gelman2017prior}.

\end{document}